\documentclass{article}
\usepackage{spconf,amsmath,graphicx}

\usepackage{xcolor,soul,framed} 

\colorlet{shadecolor}{yellow}

\usepackage{array}
\usepackage{url}

\usepackage{cite}
\usepackage{booktabs}
\usepackage{balance} 
\usepackage{multirow}
\usepackage{multicol}
\usepackage{amsfonts}
\usepackage{float}
\usepackage{bbold}

\usepackage{hyperref}
\usepackage{etoolbox,siunitx}
\robustify\bfseries
\usepackage{balance}

\usepackage{amssymb}
\usepackage{pifont}
\newcommand{\cmark}{\ding{51}}%
\usepackage{scalerel}

\title{NeuroHeed+: Improving Neuro-steered Speaker Extraction\\with Joint Auditory Attention Detection}
\name{Zexu Pan, Gordon Wichern, Fran\c{c}ois G. Germain, Sameer Khurana, Jonathan Le Roux}
\address{
  Mitsubishi Electric Research Laboratories (MERL), Cambridge, MA, USA
  }
\begin{document}
\ninept
\maketitle
\setlength{\abovedisplayskip}{4pt}
\setlength{\belowdisplayskip}{4pt}

\begin{abstract}
Neuro-steered speaker extraction aims to extract the listener's brain-attended speech signal from a multi-talker speech signal, in which the attention is derived from the cortical activity. This activity is usually recorded using electroencephalography (EEG) devices. Though promising, current methods often have a high speaker confusion error, where the interfering speaker is extracted instead of the attended speaker, degrading the listening experience. In this work, we aim to reduce the speaker confusion error in the neuro-steered speaker extraction model through a jointly fine-tuned auxiliary auditory attention detection model. The latter reinforces the consistency between the extracted target speech signal and the EEG representation, and also improves the EEG representation. Experimental results show that the proposed network significantly outperforms the baseline in terms of speaker confusion and overall signal quality in two-talker scenarios.

\end{abstract}
\begin{keywords}
Cocktail party problem, auditory attention, speaker extraction, EEG,  multi-modal
\end{keywords}
%

\section{Introduction}
\label{sec:introduction}
The human brain has a remarkable ability to focus its auditory attention on a particular stimulus, such as a target speech signal, while filtering out other stimuli, such as interfering speech signals, noise, and reverberation, a phenomenon also known as the ``cocktail party effect''~\cite{bronkhorst2000cocktail,cherry1953some}. Mimicking such ability in machines, speech separation and speaker extraction algorithms have transformed the development of hearing aids~\cite{wang2017deep} and served as important front-ends for many speech processing tasks including speech recognition~\cite{wang2022predict}, speaker localization~\cite{qian2021multi}, and speech emotion recognition~\cite{pan2020multi}.

Speech separation algorithms separate a multi-talker speech signal into individual clean streams~\cite{hershey2016deep,luo2020dual,wang2023tf}, reaching exceptional performance when the number of speakers is known. However, the separated speech signals have no association with a listener's attention. An additional algorithm is required to detect which of the separated signals is desired, using auxiliary references such as visual signals~\cite{pan2022seg} or brain signals~\cite{geravanchizadeh2021ear,o2017neural,han2019speaker,o2015attentional,van2016eeg}. The performance of such cascaded systems may be limited as each algorithm is optimized independently.

Speaker extraction algorithms are better suited to emulating the ``cocktail party effect'', as they unify the separation and detection steps into a single network to extract only the speech from a target speaker. They typically use an auxiliary reference as prior knowledge to distinguish the target speaker from interfering speakers. The most widely studied auxiliary reference is a pre-recorded speech signal~\cite{wang2019voicefilter,vzmolikova2019speakerbeam,spex_plus2020}, however, it can be cumbersome to pre-record many people's voices and select the right one to use for speaker extraction. Other auxiliary references include visual recordings such as face~\cite{ephrat2018looking,wu2019time,pan2021reentry,usev21}, gesture~\cite{pan2022seg}, direction information~\cite{elminshawi2023beamformer,tesch2023spatially} or other unique characteristics of the target speaker~\cite{tzinis2022heterogeneous}, with the limitation that it is not always feasible for the listener to visually track the target speaker, or to reliably obtain direction information.

While selecting the right auxiliary reference to use for speaker extraction is hard, an alternative way is to directly model the listener's attention through the neuronal response of their cortical activity, which reflects an interaction of external stimuli with spontaneous patterns produced endogenously~\cite{ringach2009spontaneous,harris2011cortical}. Among various cortical activity recording devices, electroencephalography (EEG) stands out as it is non-intrusive and cost-efficient with high temporal resolution. Auditory attention detection (AAD) studies~\cite{eeg_use21} show that EEG signal could be used to select one of the stimuli that the listener is focusing on in a multi-talker scenario with impressive accuracy. The stimuli studied in AAD are often the clean signals~\cite{cai2021auditory,borsdorf2023multi} or the separated signals from the speech separation algorithm~\cite{o2015attentional,van2016eeg}.

The findings in AAD studies enabled neuro-steered speaker extraction, in which the auxiliary reference is the spontaneous neuronal activity, usually recorded using EEG devices. The brain-informed speech separation (BISS) model~\cite{biss2020} utilizes the reconstructed speech envelope from the EEG signal as the auxiliary reference. The U-shaped brain-enhanced speech denoiser (UBESD) model~\cite{hosseini2022} and the brain-assisted speech enhancement network (BASEN)~\cite{zhang2023basen} go a step further and directly model the EEG signal with a temporal convolutional neural network and fuse the EEG signal with feature-wise linear modulation or convolutional multi-layer cross-attention. The neuro-steered speaker extraction (NeuroHeed)~\cite{pan2023neuroheed} is the current state-of-the-art (SOTA) model. It adopts self-attention for the EEG encoder and proposes an online auto-regressive speaker self-enrollment strategy to reinforce the speaker cue.

\begin{figure*}[t]
  \centering
  \includegraphics[height=3.3cm]{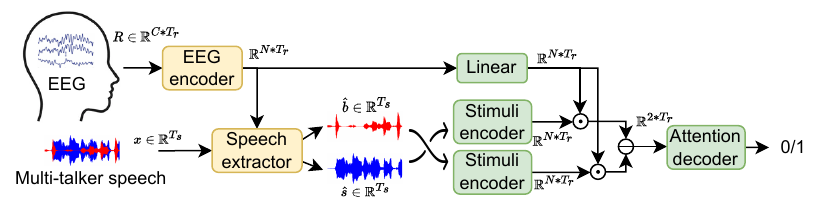}
  \vspace{-7mm}
  \caption{Our proposed NeuroHeed+ model, which jointly optimizes the speaker extraction model NeuroHeed~\cite{pan2023neuroheed} (in yellow), and an auxiliary auditory attention detection model (in green). The symbol $\ominus$ represents the concatenation of embeddings along the channel dimension, while the symbol $\odot$ refers to the inner product.}
  \vspace{-4mm}
  \label{fig:network}
\end{figure*}

Although promising, the correlation of the attended speech with the EEG signal is not as strong as compared to using the corresponding lip recording or a reference speech signal as the auxiliary reference cue. Therefore, NeuroHeed often extracts the interfering speaker instead of the attended one, which is referred to as a speaker confusion error. It creates a negative impact, especially in hearing aids, when the listener is forced to listen to the wrong speaker and cannot switch the attention back. In this work, we aim to improve NeuroHeed by reducing the speaker confusion error.

We draw inspiration from recent AAD studies~\cite{cai2021auditory,borsdorf2023multi}, which are very successful in EEG-speech association. We propose NeuroHeed+, which jointly optimizes the SOTA NeuroHeed model with an auxiliary AAD model. The AAD model maps the EEG signal and the separated speech stimuli in a common feature space, and pushes the EEG and target stimulus representations together while pulling the EEG and interfering stimulus representations away. This reinforces the consistency of the extracted target speech signal to the EEG representation, as well as improves the EEG representation. Experimental results show that the proposed network significantly outperforms NeuroHeed on speaker confusion as well as overall signal quality in two-talker scenarios on the KUL dataset~\cite{das2019auditory}.

\vspace{-.1cm}
\section{Related work: NeuroHeed}
\vspace{-.1cm}

Denote a multi-talker discrete-time speech signal as $x$, that consists of the sum of the target speech signal $s$ and the interfering speech signal $b$:
\begin{equation}
    \label{eqa:speaker_extraction}
    x = s + b  \;  \in \mathbb{R}^{T_s}
\end{equation}
where $T_s$ is the length of the speech signal. The SOTA neuro-steered speaker extraction model NeuroHeed~\cite{pan2023neuroheed} extracts an estimate of $s$ denoted $\hat{s}$, from $x$, conditioned on the EEG signal $R\in \mathbb{R}^{C*T_r}$ as the reference cue, where $C$ is the number of channels and $T_r$ is the length of the EEG signal. The different lengths $T_s$ and $T_r$ reflect the fact that the sampling rate is different between EEG and audio signals, but they are recorded over the same duration.

The NeuroHeed model is presented on the left panel of Fig.~\ref{fig:network}, and consists of an EEG encoder and a speech extractor. The original NeuroHeed only extracts $\hat{s}$, without estimating the interfering signal $\hat{b}$. NeuroHeed adopts self-attention layers~\cite{borsdorf2023multi} for its EEG encoder and a time-domain dual-path recurrent neural network (DPRNN)~\cite{luo2020dual} for its speech extractor. 

Due to the non-intrusive nature of EEG devices in capturing brain signals, the correlation of the attended speech with the EEG signal is usually weak. Generally, the reconstructed speech envelope from the EEG signal and the actual stimuli have a Pearson’s correlation of less than $0.3$~\cite{biss2020}. Therefore, to reduce speaker confusion errors in the NeuroHeed model, one needs to improve both a) the capability of an EEG encoder to extract a more discriminative EEG representation, and b) the ability of the speech extractor network to correlate the EEG representation to the target speech signal in the mixture speech signal.

\vspace{-.1cm}
\section{Proposed model: NeuroHeed+}
\vspace{-.1cm}

Our proposed NeuroHeed+ model is depicted in Fig.~\ref{fig:network}, and is an extension of the NeuroHeed model. We propose a multi-task learning framework to jointly perform speaker extraction and auditory attention detection. The speaker extraction network estimates both the target speech signal and the interfering speech signal, while the AAD model recognizes which of the speech signals correlates to the EEG signal. The two networks are separately trained first, and then cascaded and jointly fine-tuned. The AAD network is only used to better train the EEG encoder and the speech extractor, and is discarded at inference time. 

\vspace{-.1cm}
\subsection{Speaker extraction}
\vspace{-.1cm}

The speaker extraction model is shown on the left panel of Fig.~\ref{fig:network}, and consists of an EEG encoder and a speech extractor. We modify the original NeuroHeed to estimate both the target speech signal $\hat{s}$ and the interfering speech signal $\hat{b}$, such that the cascaded AAD model has access to both.

Different from speech separation models where permutation invariant training~\cite{Isik2016Interspeech09, yu2017permutation} is needed, we train the model to always estimate the target signal at a fixed output stream taking advantage of the auxiliary EEG signal. The EEG encoder and the speech extractor are optimized end-to-end using the scale-invariant signal-to-noise ratio (SI-SDR)~\cite{le2019sdr} loss function:
\begin{equation}
    \label{eqa:speech}
    \mathcal{L}_{\text{SE}} = \frac{1}{2}(\mathcal{L}_{\text{SI-SDR}}(s, \hat{s}) + \mathcal{L}_{\text{SI-SDR}}(b, \hat{b})),
\end{equation}
where
\begin{equation}
    \label{eqa:loss_sisnr}
    \mathcal{L}_{\text{SI-SDR}}(s, \hat{s}) = - 20 \log_{10} \frac{\big\|\frac{<\hat{s},s>}{\|s\|^2}s\big\|}{\big\|\hat{s} - \frac{<\hat{s},s>}{\|s\|^2}s\big\|}.
\end{equation}

\vspace{-.1cm}
\subsection{Auditory attention detection}
\vspace{-.1cm}

The AAD model is also shown in Fig.~\ref{fig:network}. It consists of an EEG encoder that is shared with the NeuroHeed model; a linear layer for EEG representation adaptation; a stimuli encoder that is formed by a time-domain speech encoder~\cite{luo2020dual} with a suitable stride size making the speech representation match the time resolution of the EEG representation, a positional encoding layer, and $5$ layers of the self-attention network; and a back-end attention decoder that is formed by two convolutional layers, one adaptive average pooling layer~\cite{cai2021auditory,borsdorf2023multi}, and a sigmoid layer.

The AAD model is first trained to discriminate which of the two clean speech signals $s$ or $b$ is the target speech that corresponds to the EEG signal. This is done by mapping the EEG signal and the two speech signals into a common feature space, and performing a dot-product operation between the EEG feature and the two speech features to obtain the respective similarity scores. The similarity scores are concatenated and passed through the attention decoder to reach a decision.

We minimize the following binary cross-entropy loss for AAD model training:
\begin{equation}
    \mathcal{L}_{\text{AAD}} = - y\log(\hat{y}) - (1-y)\log(1-\hat{y}),
\end{equation}
where $y \in \{0,1\}$ indicates which of the speech signals corresponds to the EEG signal, while $\hat{y}$ is the predicted probability.

It is worth noting that the EEG encoder is initialized from the trained NeuroHeed model and fixed during AAD model training so that it does not become inconsistent with what NeuroHeed expects. The EEG encoder is later unfrozen during the joint fine-tuning of NeuroHeed and the AAD model, so that the fine-tuning can improve the EEG representation for speaker extraction.

\vspace{-.1cm}
\subsection{Joint optimization}
\vspace{-.1cm}

After the NeuroHeed and AAD models are individually trained, they are cascaded and jointly fine-tuned as shown in Fig.~\ref{fig:network}. Instead of feeding $s$ and $b$ to the AAD model, we feed the shuffled separated signals $\hat{s}$ and $\hat{b}$ to the AAD model and train it to match $\hat{s}$ with the EEG signal as opposed to $\hat{b}$. The fine-tuning loss is defined as:
\begin{equation}
    \mathcal{L}_{\text{fine-tune}} = \mathcal{L}_{\text{SE}} + \alpha \cdot \mathcal{L}_{\text{AAD}}
    \label{eqa:finetune}
\end{equation}
where $\alpha$ is a scalar weight balancing the two tasks. 

The speaker extraction loss $\mathcal{L}_{\text{SE}}$ constrains the speech quality of the extracted signals. 
The AAD loss $\mathcal{L}_{\text{AAD}}$ backpropagates through the EEG encoder to improve the EEG representation, and also through the speaker extractor to extract discriminative speech signals from the EEG representation, improving the consistency between the separated signal $\hat{s}$ and the EEG representation.

\vspace{-.1cm}
\section{Experimental setup}
\vspace{-.1cm}
\subsection{Database}
\vspace{-.1cm}

Following~\cite{pan2023neuroheed}, we examine our proposed NeuroHeed+ model on the publicly available KULeuven (KUL) dataset~\cite{das2019auditory}. There are $16$ normal hearing subjects, with $20$ trials per subject, in which the subjects listen to concurrent speech with plugged-in earphones, and one speech signal is played in each ear. The subjects are instructed to listen to the speech in one ear while ignoring the speech in the other ear. We used the first $8$ trials for each subject, where they attend to a given signal for the first time. The speech signals are from $4$ Dutch stories spoken by $2$ male speakers. The speech signals for our speaker extraction models are sampled at $8$ kHz. The raw EEG signal is recorded using the BioSemi ActiveTwo system at $8192$ Hz with $C=64$ channels.

As in the experimental evaluation of NeuroHeed \cite{pan2023neuroheed}, we evaluate our models under a known-subject and known-speaker scenario. To do so, we randomly split each trial into training, validation, and test sets with a ratio of $75\%$, $12.5\%$, and $12.5\%$, respectively, without overlap of speech stimulus between sets. For the training set, we use mixture signal augmentation, in which the target stimulus is mixed with the interfering stimulus at a random signal-to-noise (SNR) ratio between $-10$ dB and $10$ dB. There are $400,000$, $3,000$, and $3,000$  utterances for training, validation, and test sets. 

\vspace{-.1cm}
\subsection{Hyper-parameter settings}
\vspace{-.1cm}

The hyper-parameters of the speaker extraction model exactly follow NeuroHeed~\cite{pan2023neuroheed}, with $N$ set to $64$. 
For the AAD model stimuli encoder, the time-domain speech encoder has a 1D convolutional layer (Conv1D) with input size $1$, output size $N*2$, kernel size $120$, and stride size $60$, followed by a rectified linear activation (ReLU), a layer normalization, and a linear layer with input size $N*2$ and output size $N$. The self-attention layers in the stimuli encoder have input size $N$, feedforward size $N*4$, number of heads $1$, and dropout $0.1$.
For the AAD model attention decoder, the first Conv1D has input and output sizes $2$, kernel size $15$, stride $7$, while the second Conv1D has input size $2$ and output size $1$, kernel size $15$, stride $7$, and a parametric ReLU is used between the two convolutional layers.

\vspace{-.1cm}
\subsection{Training details}
\vspace{-.1cm}

We use PyTorch to conduct our experiments. All models are trained on $2$ GPUs with \SI{48}{\giga\byte} RAM each. The Adam optimizer is used with a learning rate (lr) warm-up as follows for the first $15,000$ training steps:
\begin{equation}
    \text{lr}(n) = 0.1 \cdot N^{-0.5} \cdot n \cdot 15,000^{-1.5}
\end{equation}
where $n$ is the step number. After the warm-up is done, the lr is halved when the best validation loss (BVL) does not improve within 6 consecutive epochs. The training stops when the BVL does not improve for 10 subsequent epochs. For the joint fine-tuning, the model weights are initialized from the previously trained weights, but the optimizer and lr scheduler are re-initialized. 
We share a common model between subjects, which has experimentally shown to have better performance than subject-specific models.

\vspace{-.1cm}
\subsection{Evaluation metrics}
\vspace{-.1cm}

We use the improvement in SI-SDR (SI-SDRi) and the improvement in signal-to-distortion ratio (SDRi) to evaluate the signal quality, while using the improvement in perceptual evaluation of speech quality (PESQi) and the improvement in short term objective intelligibility (STOIi) to evaluate the perceptual quality and intelligibility of the extracted speech with respect to the unprocessed multi-talker speech signals. To evaluate the speaker confusion error, we report the percentage positive rate (PPR)~\cite{pan2023neuroheed}, which is defined as the percentage of extracted speech signals that satisfy both i) a positive SI-SDRi value and ii) a higher SI-SDRi value with respect to the target speech signal than to the interfering speech signal. The higher the better for all metrics.

\vspace{-.1cm}
\section{Results}
\vspace{-.1cm}

We first present in Sec.~\ref{sec:result_tune} the results on the validation set of our proposed NeuroHeed+ with various training strategies, to select our best model. We then compare our proposed NeuroHeed+ with various baselines on the test set in~Sec~\ref{sec:result_benchmark}, to show the superiority of our model.

\begin{table}
    \centering
    \sisetup{
    detect-weight, 
    mode=text, 
    tight-spacing=true,
    round-mode=places,
    round-precision=1,
    table-format=2.1
    }
    \caption{Validation set results for NeuroHeed+ with various configurations. We find our best model according to the reported SI-SDR value in dB. We give a system number (Sys.\ \#) to every different system. $\alpha$ is the scalar weight for the fine-tuning loss. We initialize (Init) and fix different modules during the joint training stage, such as the EEG encoder (EE), speaker extractor (SE), and the green modules in the AAD module (AAD) in Fig.~\ref{fig:network}. }
    \addtolength{\tabcolsep}{-3pt}
    \resizebox{\linewidth}{!}{
    \begin{tabular}{c S[table-format=1.1] *{5}{c} S} 
       \toprule
        {Sys.\ \#}     &~~~~$\alpha$~~~~   &\!\!\!\!Init SE\&EE    &Init AAD   &Fix AAD   &Fix EE  &Fix SE   &{SI-SDR}\\
        \midrule  
        0~\cite{pan2023neuroheed}\!\!\!\!\!\!\!\!\!\!\!\!   &{-}    &- &- &- &- &- &13.42\\
        1   &0.0    &- &- &- &- &- &12.341\\
        2   &1.0    &- &- &- &- &- &13.18 \\
        3   &1.0    &\cmark &- &- &- &- &14.019\\
        4   &1.0    &\cmark &\cmark &- &- &- &\bfseries14.254\\
        5   &1.0    &\cmark &\cmark &\cmark &- &- &14.160\\
        6   &1.0    &\cmark &\cmark &\cmark &\cmark &- &13.996\\
        7   &1.0    &\cmark &\cmark &\cmark &- &\cmark  &13.67\\
        \bottomrule
    \end{tabular}
    }
    \addtolength{\tabcolsep}{3pt}
    \vspace*{-5mm}
    \label{tab:val_init}
\end{table}

\vspace{-.1cm}
\subsection{Model tuning}
\label{sec:result_tune}
\vspace{-.1cm}

In Table~\ref{tab:val_init}, 
Sys.\ 0 is the original NeuroHeed model~\cite{pan2023neuroheed}, which obtains $13.4$ dB. Sys.\ 1 is the modified NeuroHeed model that estimates both $\hat{s}$ and $\hat{b}$, without joint training. The SI-SDR drops by $1.1$ dB from Sys.\ 0 to Sys.\ 1, this is probably because the network has limited capacity, thus the performance drops when estimating both signals. 
Sys.\ 2 is the NeuroHeed+ model with all modules trained from scratch. The joint learning improves the SI-SDR by $0.9$ dB compared with Sys.\ 1, but is still not better than Sys.\ 0.

We next explore various initialization and fine-tuning approaches. 
In Sys.\ 3, the Speaker extractor and EEG encoder are initialized from Sys.\ 1  and fine-tuned with Eq.~(\ref{eqa:finetune}), improving the SI-SDR to $14.0$ dB. 
In Sys.\ 4, the AAD modules are also initialized from pre-training before fine-tuning the whole system, and the SI-SDR further improves to $14.3$ dB. 
In Sys.\ 5, the AAD modules are fixed, and we fine-tune the EEG encoder and speaker extractor, obtaining a similar SI-SDR to Sys.\ 4. 
In Sys.\ 6, we further fix the EEG encoder and only fine-tune the speaker extractor, leading to a degradation of the SI-SDR by $0.3$ dB compared to Sys.\ 4. In other words, the $0.8$ dB gain from Sys.\ 2 to Sys.\ 6 represents the improvements in the speaker extractor's ability brought by the auxiliary AAD task, to correlate the EEG representations with the target speaker during the extraction process.
In Sys.\ 7, we only fine-tune the EEG encoder, and the SI-SDR degrades by $0.6$ dB from our best model Sys.\ 4. In other words, the $0.5$ dB gain from Sys.\ 2 to Sys.\ 7 represents the improvements in the EEG representation brought by the auxiliary AAD task.

We note that the proposed auxiliary AAD task may be sub-optimal during joint training, as it leaves the opportunity for the speech extractor to learn to output $\hat{s}$ with some EEG information encoded implicitly and for the stimuli encoder to learn to decode that EEG information. However, we still see improvements in the speaker extraction task, which could be because the SI-SDR loss $\mathcal{L}_{\text{SE}}$ constrains $\hat{s}$ and regularizes the speaker extractor training. In addition, the initialization enables the modules to start from a better state instead of quickly converging to a sub-optimal solution.

\begin{table}
    \centering
    \sisetup{
    detect-weight, 
    mode=text, 
    tight-spacing=true,
    round-mode=places,
    round-precision=1,
    table-format=2.1
    }
    \caption{Validation set results for NeuroHeed+ with various scalar weight $\alpha$ for the fine-tuning loss. We find our best model according to the reported SI-SDR value in dB. }
    \addtolength{\tabcolsep}{5pt}
    \resizebox{.5\linewidth}{!}{
    \begin{tabular}{c c S} 
       \toprule
        {Sys.\ \#}     &$\alpha$    &{SI-SDR}\\
        \midrule  
        8   &0.001     &13.93\\
        9   &0.01      &14.14\\
        10   &0.1       &14.06\\
        4   &1       &\bfseries14.254\\
        11   &10     &13.87\\
        12   &100    &13.24\\
        \bottomrule
    \end{tabular}
    }
    \addtolength{\tabcolsep}{5pt}
    \vspace*{-5mm}
    \label{tab:val_alpha}
\end{table}

In Table~\ref{tab:val_alpha}, we present validation set results for the NeuroHeed+ model with various $\alpha$, using the same initialization and fine-tuning strategy as Sys.\ 4. We can see that the best $\alpha$ value is $1$. Therefore, we select Sys.\ 4 as our final best NeuroHeed+ model.

\vspace{-.1cm}
\subsection{Comparison with baselines}
\label{sec:result_benchmark}
\vspace{-.1cm}

In Table~\ref{tab:test}, we compare the test set results of NeuroHeed+ with those of various baselines.
Sys.\ 13 is an oracle system for upper-bound analysis, which performs speech separation using DPRNN~\cite{luo2020dual} first, and then performs ground-truth association by selecting the separated speech signal that has the highest SI-SDR with $s$. Therefore, it has a PPR of $100\%$, and a very high SI-SDRi value of $19.4$ dB.
Sys.\ 14 is a baseline that performs speech separation using DPRNN first, and then performs the association with a separately trained AAD network. 
Sys.\ 15 has the same pipeline as Sys.\ 14 except that the DPRNN is fine-tuned together with the AAD network. 
We can see that our proposed NeuroHeed+ model (Sys.\ 4) outperforms by a large margin in terms of all metrics both Sys.\ 14 and 15, which perform speech separation without EEG input instead of target speech extraction.
The proposed NeuroHeed+ also outperforms NeuroHeed by $1.3$ dB in SI-SDRi, $1.2$ dB in SDRi, $0.13$ in PESQi, $0.02$ in STOIi, and $1.5\%$ in PPR (a $16\%$ relative error reduction).

\begin{figure}[t]
\begin{minipage}[t]{.49\linewidth}
  \centering
  \centerline{\includegraphics[width=\linewidth]{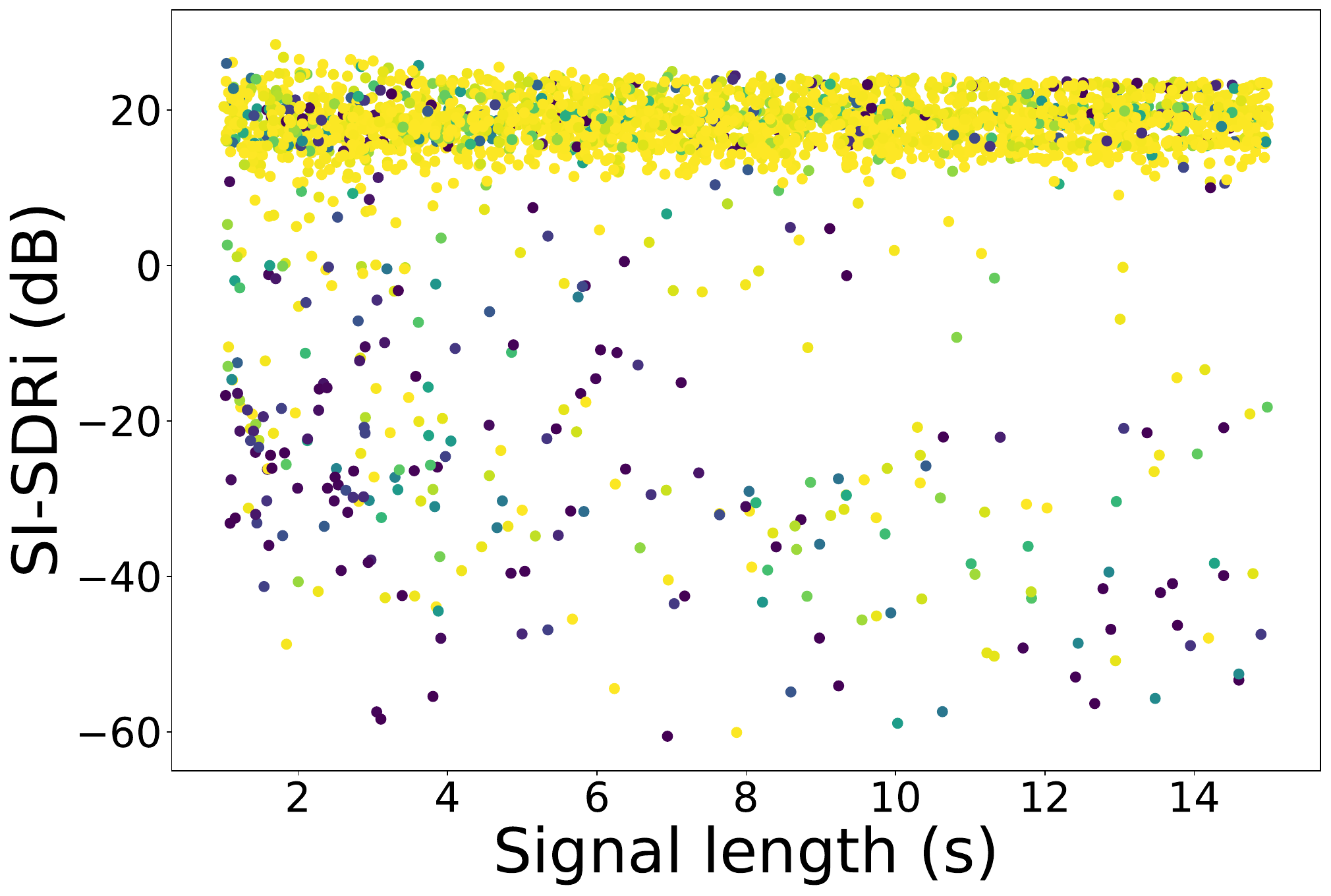}}
  \centerline{\scalebox{0.8}{(a) Sys.\ 0: NeuroHeed}}\medskip
\end{minipage}
\hfill
\begin{minipage}[t]{.49\linewidth}
  \centering
  \centerline{\includegraphics[width=\linewidth]{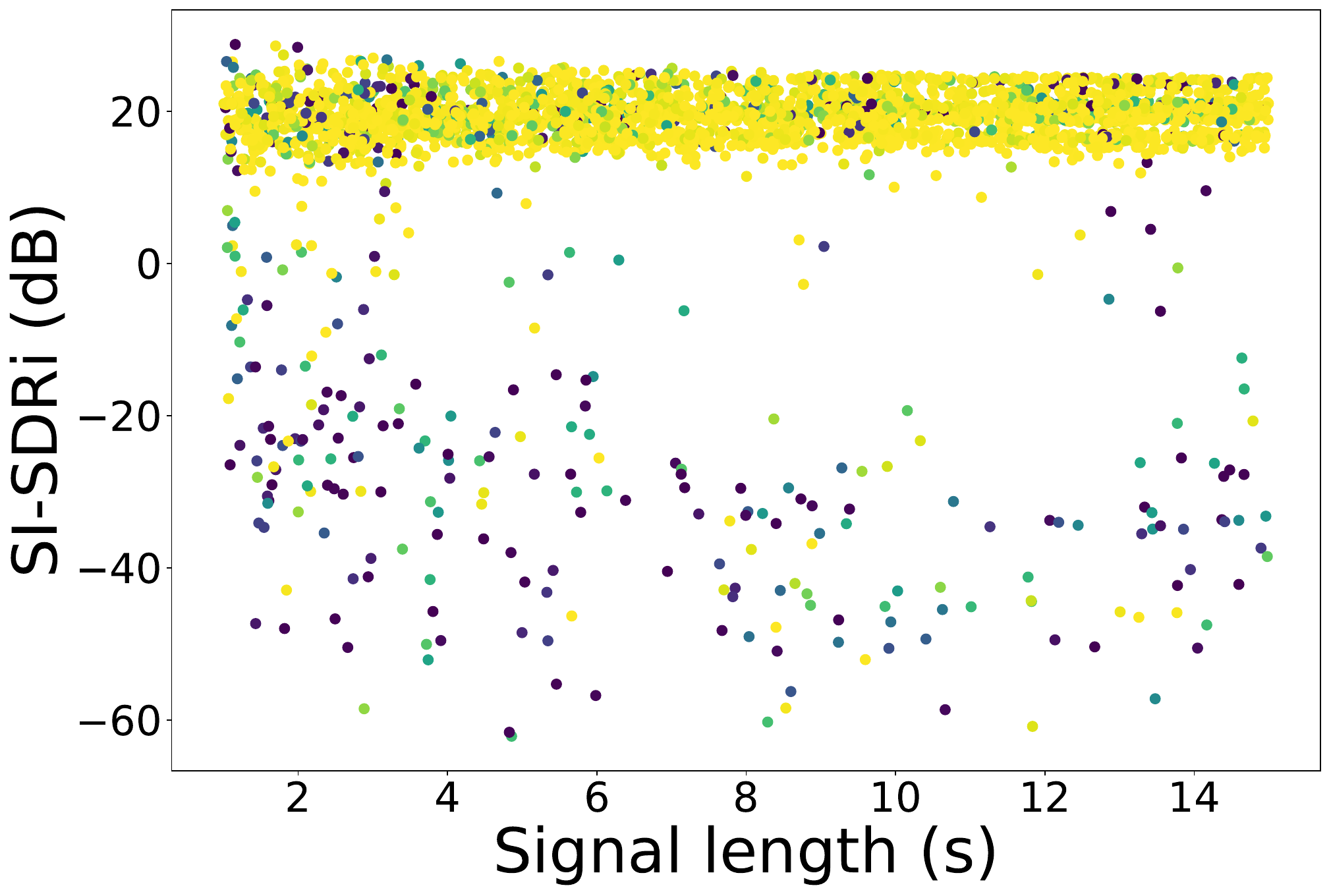}}
  \centerline{\scalebox{0.8}{(b) Sys.\ 4: NeuroHeed+}}\medskip
\end{minipage}
\vspace{-2mm}
\caption{SI-SDRi scatter plot of extracted speech signal for various lengths of audio signals in the test set, by (a) the NeuroHeed model~\cite{pan2023neuroheed}, and (b) our proposed NeuroHeed+ model. The color represents the AAD probability of making a correct attention detection, with bright yellow meaning correct and dark purple incorrect.}
\vspace{-5mm}
\label{fig:histogram}
\end{figure}

Fig.~\ref{fig:histogram} presents the scatter plot of SI-SDRi of the extracted speech signals for signal lengths ranging from $1$ s to $15$ s. For both (a) NeuroHeed and (b) NeuroHeed+, the majority of samples have SI-SDRi around $20$ dB, meaning that the models extract the correct target speaker with high signal quality. 
As the signal length increases, both models have fewer samples having negative SI-SDRi values, meaning that the model is able to learn from the longer context when a longer EEG signal is available.
Overall, NeuroHeed+ has fewer low SI-SDRi samples compared with NeuroHeed, meaning that NeuroHeed+ makes fewer speaker confusion errors, which explains the average $1.3$ dB SI-SDRi gain shown in Table~\ref{tab:test}. 

The samples in Fig.~\ref{fig:histogram} are plotted with colors, which represent the AAD output probability score of correctly associating the EEG representation with the clean target signal $s$ instead of the interfering signal $b$, with bright yellow meaning correct and dark purple 
incorrect. Because the color is associated with how correlated the EEG representation is to $s$, it gives an indication as to how easy speaker extraction is. However, the speech extractor uses the EEG representation as conditioning to extract part of a mixture speech signal, 
so they may not always agree, with one succeeding in selecting the correct speaker while the other fails.
Our proposed joint training aims to promote agreement between the speaker extraction model and the AAD model, meaning in particular that for samples where AAD makes accurate classification, the speaker extraction model will make fewer speaker confusion errors. As shown in Fig.~\ref{fig:histogram}, NeuroHeed+ indeed has fewer low SI-SDRi samples with a bright yellow color compared with NeuroHeed, meaning that our proposed joint training is able to improve the speaker extraction model in agreeing with the AAD model on those samples.

\begin{table}
    \centering
    \sisetup{
    detect-weight, 
    mode=text, 
    tight-spacing=true,
    round-mode=places,
    round-precision=1,
    table-format=2.1
    }
    \caption{Test set results for EEG-steered speaker extraction models. SI-SDRi and SDRi are reported in dB. The percentage positive rate (PPR) is the percentage of the extracted speech in the test set that has both a positive SI-SDRi value and a higher SI-SDRi value with respect to the attended speech than to the interfering speech. The higher the PPR, the lower the speaker confusion error.}
    \addtolength{\tabcolsep}{-1.5pt}
    \resizebox{\linewidth}{!}{
    \begin{tabular}{*{1}{S[table-format=1,round-precision=0]} l SS *{2}{S[round-precision=2,table-format=1.2]} S} 
       \toprule
        {Sys.\ \#}     &Model  &{SI-SDRi}   &{SDRi}   &{PESQi}  &{STOIi} & {PPR}\\
        \midrule  
        13   &Separation-PIT~\cite{luo2020dual} 
            &19.4  &19.6  &1.22025	&0.230856  &100.0 \\  
        14   &Separation-AAD 
            &4.6394	&9.168	&0.64253	&0.01904	&74.7 \\
        15   &\;+ jointly fine-tuned 
            &12.8982	&15.1787	&0.95483	&0.13577	&88.5 \\
        \midrule
        16   &BISS~\cite{biss2020}
            &-0.1	&0.5	&-0.08	&-0.03	&59.4\\
        17   &UBESD~\cite{hosseini2022}
            &5.1   &5.8   &0.09  &0.03  &80.9\\
        18   &BASEN~\cite{zhang2023basen}
            &5.6   &6.7   &0.22  &0.03  &75.6\\
        0   &NeuroHeed~\cite{pan2023neuroheed}
            &14.3  &15.5  &0.95  &0.16  &90.8 \\
        \midrule
        4   &\bfseries{NeuroHeed+}
            &\bfseries15.554	&\bfseries16.71488	&\bfseries1.0845	&\bfseries0.17656	&\bfseries92.2666\\
        \bottomrule
    \end{tabular}
    }
    \addtolength{\tabcolsep}{1.5pt}
    \vspace*{-3mm}
    \label{tab:test}
\end{table}

\vspace{-.1cm}
\section{Conclusion}
\vspace{-.1cm}

In this work, we reduce the speaker confusion error for the SOTA neuro-steered speaker extraction model NeuroHeed. We propose NeuroHeed+, which has a joint learning framework such that the speaker extraction model benefits from the auxiliary AAD task in improving the EEG representation, and improving the EEG-speech association in the speaker extraction processes. Experimental results show that the proposed NeuroHeed+ is effective in extracting the correct speakers and achieving a new SOTA on the KUL dataset.

\newpage


\footnotesize
\bibliographystyle{IEEEbib}
\bibliography{IEEEabrv,Bibliography}

\end{document}